\documentclass[a4paper,12pt]{article}

\usepackage{amsmath,amssymb,amsfonts}
\usepackage{epsfig,cite, cancel, slashed,cite}
\usepackage[footnotesize]{caption}
\usepackage{longtable, color}

\pdfoutput=1

\newlength{\xtrawidth}
\newlength{\xtraheight}
\newcommand{\sref}[1]{Subsection~\ref{#1}}

\newcommand{\cref}[1]{Chapter~\ref{#1}}
\newcommand{\bcenter}{\begin{center}}
\newcommand{\ecenter}{\end{center}}
\newcommand{\beq}{\begin{equation}}
\newcommand{\eeq}{\end{equation}}
\newcommand{\bea}{\begin{eqnarray}}
\newcommand{\eea}{\end{eqnarray}}
\newcommand{\bean}{\begin{eqnarray*}}
\newcommand{\eean}{\end{eqnarray*}}
\newcommand{\ba}{\begin{array}}
\newcommand{\ea}{\end{array}}
\newcommand{\ben}{\begin{enumerate}}
\newcommand{\een}{\end{enumerate}}
\newcommand{\bi}{\begin{itemize}}
\newcommand{\ei}{\end{itemize}}
\newcommand{\bd}{\begin{description}}
\newcommand{\ed}{\end{description}}
\newcommand{\bdiag}{\begin{diagram}}
\newcommand{\ediag}{\end{diagram}}
\def\fnote#1#2{\begingroup\def\thefootnote{#1}\footnote{#2}
     \addtocounter{footnote}{-1}\endgroup}

\def\IR{\mathbb{R}}
\def\IP{\mathbb{P}}
\def\IZ{\mathbb{Z}}
\def\cO{{\mathcal O}}
\def\cK{{\mathcal K}}

\newcommand{\sseq}[5]{0 \to #1 \stackrel{#4}{\longrightarrow} #2  \stackrel{#5}{\longrightarrow} #3 \to 0}
\newcommand{\nn}{\nonumber}

\newcommand{\cA}{{\cal A}}

\newcommand\cM{{\cal M}}

\newcommand{\be}{\begin{equation}}
\newcommand{\ee}{\end{equation}}

\newcommand{\comment}[1]{}

\def\IR{\mathbb{R}}
\def\IP{\mathbb{P}}
\def\IZ{\mathbb{Z}}

\numberwithin{equation}{section}

\begin{document}


\title{\LARGE \bf{Heterotic Bundles on Calabi-Yau Manifolds with Small Picard Number }}
\author{
Yang-Hui He${}^1$,
Maximilian Kreuzer,  
Seung-Joo Lee${}^2$,
Andre Lukas${}^3$
}
\date{}
\maketitle
\begin{center}
{\small
${}^1${\it
Department of Mathematics, City University, London,\\ 
Northampton Square, London EC1V 0HB, U.K.;\\ 
School of Physics, NanKai University, Tianjin, 300071, P.R.~China;\\ 
Merton College, University of Oxford, OX14JD, U.K.}\\[0.2cm]
${}^2${\it School of Physics, Korea Institute for Advanced Study, Seoul 130-722, Korea}\\[0.2cm]
${}^3${\it Rudolf Peierls Centre for Theoretical Physics, Oxford
  University,\\
$~$ 1 Keble Road, Oxford, OX1 3NP, U.K.}\\[0.8cm]
{\bf In memoriam Maximilian Kreuzer}\\
\fnote{}{hey@maths.ox.ac.uk}
\fnote{}{s.lee@kias.re.kr}
\fnote{}{lukas@physics.ox.ac.uk}
}
\end{center}
\abstract{
We undertake a systematic scan of vector bundles over spaces from the largest database of known Calabi-Yau three-folds, in the context of heterotic string compactification. Specifically, we construct positive rank five monad bundles over Calabi-Yau hypersurfaces in toric varieties, with the number of K\"ahler moduli equal to one, two, and three and extract physically interesting models. We select models which can lead to three families of matter after dividing by a freely-acting discrete symmetry and including Wilson lines. About {2000} such models on {two} manifolds are found. 
}

\newpage 
\tableofcontents
\section{Introduction}
Algorithmic and computational methods in string compactifications provide a valuable set of tools in the quest to connect string theory with particle physics. The subject of constructing vacua of superstring theory, notably large databases of Calabi-Yau manifolds, has been an ongoing enterprise since the first seminal work on string phenomenology \cite{Candelas:1985en}.
The largest set to date is that compiled by Kreuzer and Skarke~\cite{Kreuzer:1995cd,Avram:1997rs,Kreuzer:2000qv,Kreuzer:2000xy,Kreuzer:2002uu} over more than a decade, and amounts to an impressive list of some 500 million smooth Calabi-Yau three-folds.
It is, however, only with the recent advances in computational algebraic geometry that we can begin to address some of the questions related to this data set.

Over the past five years, considerable work \cite{Anderson:2011ns,Anderson:2008uw,Anderson:2009mh,Anderson:2007nc}, has been done on the smaller but illustrative dataset of complete intersection Calabi-Yau manifolds (CICYs)~\cite{Candelas:1987kf} (cf.~also the plethora of stable bundles over elliptically fibred Calabi-Yau three-folds in~\cite{Gabella:2008id}). Over these manifolds, various types of vector bundles such as monad bundles, extension bundles and simple sums of line bundles have been constructed in a search for physically promising models. Recently, about 200 models with the MSSM spectrum have been found in this way~\cite{Anderson:2011ns} in a scan over about $10^{12}$ initial bundles. This should be compared with the relatively small number of heterotic standard models on Calabi-Yau manifolds known before~\cite{Braun:2005nv,Bouchard:2005ag,Anderson:2009mh}.

Extending this work to the Kreuzer-Skarke list of hypersurfaces in toric four-folds, to which we henceforth refer as ``toric Calabi-Yau three-folds'' is a natural step and an ambitious task in itself. In Ref.~\cite{He:2009wi}, we initiated this study by investigating a special class of the entire list, namely the 101 three-folds whose ambient toric varieties are smooth and, in addition, have simplicial K\"ahler cones.
This is a good starting point since many of the technicalities simplify. For example, the K\"ahler cone of such a Calabi-Yau three-fold $X$ with Hodge numbers $(h^{1,1}(X),h^{2,1}(X))$ can be brought to the positive orthant of $\IR^{h^{1,1}(X)}$, thereby simplifying the check for ``positive" line bundles. Subsequently we constructed positive monad bundles~\footnote{\sref{3.1.1} will make clear what precisely we mean by a positive monads.}, that is monads built from positive line bundles, over these spaces. We were able to find that only 11 of these manifolds admit  such positive monad bundles, with a total of about 2000 bundles, only 21 of which could possibly allow three generations of quarks and leptons.
Relaxing the positivity constraint gave 280 models which could then be candidates for model building.

In this paper, we take a more systematic approach. Building upon the experience gained from Ref.~\cite{He:2009wi} we start analysing the complete toric list, proceeding level by level in the Picard number $h^{1,1}(X)$. At low Picard number, going through the database, there are, respectively, 5, 36 and 244 manifolds with $h^{1,1}(X)$ equal to $1$, $2$ and $3$. This is a sufficiently large set for us to investigate presently.
Indeed, Calabi-Yau manifolds with small Hodge numbers have recently become an active area of research \cite{Candelas:2007ac,Braun:2011hd,Candelas:2008wb,Braun:2009qy}. Now, we are confronted with the possibility of singular ambient spaces and thus have the need to de-singularize, a technical detail which we will address, especially with the aid of the computer package PALP \cite{Kreuzer:2002uu,Braun:2011ik}. 

Let us briefly summarize the search criteria we impose in the context of heterotic compactification. We are looking for the following properties:
\begin{itemize}
\item The toric Calabi-Yau manifolds $X$ should allow for a freely-acting discrete symmetry $\Gamma$. In practice, we test for such symmetries by checking divisibility of various topological invariants of $X$, including the Euler number.
\item A stable holomorphic vector bundle $V$ over the toric Calabi-Yau manifolds $X$ with structure group $SU(N)$, so that, in particular, $c_1(V)=0$. This structure group, embedded into the ``observable" $E_8$ gauge theory, leads to an ${\cal N}=1$ four-dimensional GUT theory with gauge group $E_6$, $SO(10)$ and $SU(5)$, respectively for $N=3, 4~\text{and}~5$.
\item The second Chern class of $V$, $c_2(V)$, is constrained by the second Chern class $c_2(TX)$ of the manifold $X$ via Green-Schwarz anomaly cancellation.
\item Stability of $V$ implies that the cohomology groups $H^0(X,V)$ and $H^3(X,V)$ vanish, and, hence, the index $\mbox{ind}(V) = \frac{1}{2} \int_X c_3(V) = -h^1(X,V) + h^2(X,V)$ provides the net number of generations. We require this index to be equal to $-3|\Gamma|$, where $|\Gamma|$ is the order of a freely acting discrete symmetry $\Gamma$ on $X$. Upon taking the quotient by $\Gamma$ and breaking the GUT group with Wilson lines this leads to three net generations ``downstairs".
\item We require the absence of anti-generations. It turns out, for positive monads this condition is automatically satisfied. 
\end{itemize}

Given these mathematical conditions on the manifolds and the bundles over them, our task is clear. We need to classify all positive monad bundles over the available manifolds and extract the cases with the above properties. The paper is organized as follows. In the following section we describe the toric Calabi-Yau manifolds and some of their properties. In Section 3 we explain the general monad bundle construction, our scanning results for manifolds with $h^{1,1}(X)=1,2,3$ and a few explicit examples taken from the set we have found. The full data set of all {2000} bundles is available from the webpage~\cite{link}. Conclusions and future directions are discussed in Section 4. 

\section{The toric Calabi-Yau three-folds}
In the case of smooth toric ambient spaces discussed in Ref.~\cite{He:2009wi}, each of the associated four-dimensional reflexive polytopes provided a single family of Calabi-Yau three-folds. However, when the ambient space is singular, one may get several families since different blow-ups of the toric four-fold in general give rise to different Calabi-Yau hypersurfaces. 
Given a four-dimensional reflexive polytope, one can look for the so-called ``maximal projective crepant partial'' desingularisations(MPCP-desingularisations) of the ambient toric fourfold, which correspond to MPCP-triangulations of the (dual) polytope.\footnote{A triangulation is said to be {\it maximal} if all lattice points of the polytope are involved, {\it projective} if the K\"{a}hler cone of the resolved manifold has a nonempty interior, {\it crepant} if no points outside of the polytope are taken. In practice, all possible MPCP-triangulations of a given reflexive polytope are found by using the computer package PALP \cite{Kreuzer:2002uu}.}
They are called ``partial'' desingularisations because the singularities of the ambient space may not be fully resolved.  
Nonetheless, the Calabi-Yau hypersurface is smooth everywhere~\cite{Batyrev:1994hm} while the ambient space is only partially resolved and may still have a singular locus.  
Therefore, in order to resolve the possible singularities of Calabi-Yau three-fold, we look for all MPCP-desingularisations of the ambient space or, equivalently, all possible MPCP-triangulations of the (dual) polytope. 

\subsection{The database of CYs with small $h^{1,1}(X)$}\label{2.1}
At $h^{1,1}(X) = 1$, there are 5 manifolds $X_{1,i} ~(i=1, \cdots 5)$, presented in Table \ref{t:pic1}.
These are the cyclic manifolds and proving bundle stability is a relatively straightforward matter on such spaces~\cite{Anderson:2007nc, Anderson:2008uw}. It turns out that each of the five cyclic manifolds $X$ only admits a single MPCP-desingularisation.\footnote{For the smooth ambient cases we do not triangulate the polytopes. However, we will consider this as a single, trivial triangulation.} 
In the list is, of course, the famous quintic three-fold($X_{1,2}$) in $\IP^4$, which also appears in the CICY list \cite{Anderson:2008uw} as well as the list for toric Calabi-Yau manifolds with smooth ambient space~\cite{He:2009wi}.
The first one, $X_{1,1}$, turns out to be the quintic manifold quotiented by the freely acting Abelian (toric) $\IZ_5$ symmetry (recall that there are two freely acting $\IZ_5$--actions on the quintic; one is a toric symmetry and the other a permutation symmetry).
In Table~\ref{t:pic1}, we include the Hodge numbers and the Euler numbers for reference.
We also present the defining dual polytopes of $X$.
The dual polytopes are expressed, for compactness, by their Newton polynomials, each monomial therein being a product of the four dummy variables, $u,v,w$ and $z$, raised to the powers given by the coordinates of the corresponding vertex.\footnote{Of course, we could as well define similar Newton polynomials for the original polytopes. In this work, however, we will mainly use the dual polytopes since they are the ones directly related to the combinatorics of normal fans. So the toric data will be expressed in terms of dual polytopes.}
For example, the dual polytope for the quintic in $\IP^4$ has the 5 vertices, $(1, 0, 0, 0)$, $(0, 1, 0, 0)$, $(0, 0, 1, 0)$, $(0, 0, 0, 1)$, and $(-1, -1, -1, -1)$. 
These then give rise to the 5 monomials $u$, $v$, $w$, $z$ and $u^{-1}v^{-1}w^{-1}z^{-1}$ respectively, which constitute the 5 terms for the Newton polynomial, as can be seen from the second row of Table~\ref{t:pic1}. The coefficients are always chosen to be $+1$. For reference and completeness, we include this data, together with the analogous data for the $h^{1,1}(X)=2,3$ manifolds, on the webpage \cite{link}.
\begin{table}[h!t!]
{\renewcommand{\arraystretch}{1.1}
\[
\begin{array}{|c|c|c|c|c|}
\hline
\mbox{Manifold} & h^{1,1} & h^{2,1} & \chi & \mbox{Newton Polynomial} \\
\hline \hline
X_{1,1} & 1&21&-40 &
u v^2 w^3 z^5+u+v+w+\frac{1}{u^2 v^3 w^4 z^5} \\
X_{1,2} & 1&101&-200 &
u+v+w+z+\frac{1}{v w z u} \\
X_{1,3} & 1&103&-204 &
u+v+w+z+\frac{1}{v w z u^2} \\
X_{1,4} & 1&145&-288 &
u+v+w+z+\frac{1}{v^2 w z u^5} \\
X_{1,5} & 1&149&-296 &
u+v+w+z+\frac{1}{v w z u^4}  \\[1mm] 
\hline
\end{array}
\]}
\begin{center}
\parbox{6.5in}{\caption{{\em \sf The 5 cyclic (i.e., $h^{1,1}(X)=1$) hypersurface Calabi-Yau three-folds in toric four-folds. $X_{1,2}$ is the famous quintic. For each manifold, the Hodge numbers $h^{1,1}$ and $h^{2,1}$, the Euler number $\chi$, as well the Newton polynomial, an equivalent representation of the defining dual polytope, are given.}}
\label{t:pic1}}\end{center}
\end{table}

At $h^{1,1}(X) = 2$, there are 36 manifolds $X_{2,i}$ ($i=1,\cdots, 36$); their Hodge numbers, Euler numbers and toric data can be found in \cite{link}.
It turns out, upon searching for the MPCP-desingularisations of the ambient spaces, that 24 of them admit a unique blowup and hence give rise to a single smooth CY hypersurface each, while the remaining 12 manifolds admit two different blowups. 
For the latter 12 cases, one can compute the intersection rings of the resolved Calabi-Yau manifolds~\cite{Kreuzer:2002uu}, and the two different triangulations for a given reflexive polytope may in general give rise to two different intersection rings. 
If that happens one must consider them as two different manifolds, corresponding to two different {\it phases} of the ambient space. Otherwise,  both desingularisations correspond to the same Calabi-Yau manifold. It turns out that 3 of the 12 manifolds are of the former type and hence give rise to 6 smooth Calabi-Yau manifolds, while the other 9 are of the latter type, each providing one smooth Calabi-Yau. 
Therefore the 36 reflexive four-polytopes result in 39 smooth Calabi-Yau hypersurfaces $X$ with $h^{1,1}(X)=2$. 
These include the bi-degree $(3,3)$ hypersurface in $\IP^2 \times \IP^2$ as well as the bi-degree $(2,4)$ hypersurface in $\IP^1 \times \IP^3$; these again have been studied in Ref.~\cite{Anderson:2008uw,He:2009wi}.
In total, 9 of the 39 manifolds (including, of course, the two CICYs) have smooth, simple ambient spaces and thus have already been studied in Ref.~\cite{He:2009wi}.

At $h^{1,1}(X) = 3$, there are 244 manifolds $X_{3,i}~(i=1,\cdots, 244)$, the list of which is again given in \cite{link}. 
Of these, the tri-degree $(2,2,3)$ hypersurface in $\IP^1 \times \IP^1 \times \IP^2$ has already been probed for positive monad bundles \cite{Anderson:2008uw}. In total, 28 of these 244 manifolds (including the $(2,2,3)$ CICY) have smooth, simple ambient spaces and hence do not require desingularisations. Starting from the total of 244 manifolds including singular ones, the partial resolutions of singularities give rise to 307 smooth Calabi-Yau three-folds, 266 of which have a simplicial K\"{a}hler cone.  

In this work, we do not attempt to deal with the manifolds with $h^{1,1}(X) \geq 4$.
However, let us move on two steps further to give an idea of the numbers of available Calabi-Yau three-folds. 
At $h^{1,1}(X) = 4$, there are 1197 manifolds $X_{4,i}~(i=1,\cdots, 1197)$.
This includes the tetra-quadric $(2,2,2,2)$ hypersurface in $\IP^1 \times \IP^1 \times \IP^1 \times \IP^1$. A total of 44 have smooth, simple ambient spaces.
At $h^{1,1}(X) = 5$, there are 4990 manifolds $X_{5,i}~(i=1,\cdots, 4990)$ and only a total of 18 have smooth, simple ambient spaces, which have thus been studied already.

We can readily obtain all the Hodge numbers of the above manifolds directly from the defining (dual) polytopes.
It is important to emphasize that, unlike for previously studied classes, not all of these manifolds are {\it favourable} in the sense that all K\"ahler classes of $X$ descend from those of the ambient toric fourfold $\cA$. In other words, for non-favourable cases ${\rm Pic}(X)\simeq H^{1,1}(X, \IZ)$ is not isomorphic to  ${\rm Pic}(\cA) \simeq H^{1,1}(\cA, \IZ) \simeq \IZ^{k-4}$, where $k$ is the number of edges of the fan after a particular MPCP-triangulation. Indeed, we can check that in some cases, $h^{1,1}(X)$ of the hypersurfaces is not equal to $h^{1,1}(\cA)$ of the ambient spaces. However, for the manifolds satisfying $h^{1,1}(X) \leq 3$, there is only a single non-favourable example which arises at $h^{1,1}(X)=3$. For simplicity, this manifold will be discarded in the subsequent scans. 

\subsection{Triangulation and smoothing} \label{palp-output}
As explained in the beginning of the section, the ambient space $\cA$ is not smooth in general. Therefore, the MPCP-desingularisations procedure is required, in order to ensure the smoothness of the Calabi-Yau hypersurface $X$. Let us demonstrate this procedure with an example. Take the toric ambient space for $X_{2,4}$ from \cite{link}. The Newton polynomial is $\frac{w z u^3}{v}+\frac{u}{v}+v+w+z+\frac{v}{w z u^2}$, representing the dual polytope with the following 6 vertices in $\IZ^4$:
\begin{equation}\label{polytope24}
{\scriptsize
\left[
\begin{array}{rrrrrrl}
 0 & 3 & 0 & 0 & 1 & -2 \\
 0 & -1 & 1 & 0 & -1 & 1 \\
 1 & 1 & 0 & 0 & 0 & -1 \\
 0 & 1 & 0 & 1 & 0 & -1
\end{array}
\right]
} \ .
\end{equation}
Using the computer package PALP \cite{Kreuzer:2002uu,Braun:2011ik}, we readily find that there are two different MPCP-triangulations of this polytope, each consisting of 8 refined simplicial cones:
\begin{equation}\begin{array}{l}  \nn
111100,101101, 100111, 110110, 001111, 011110, 101011, 111010 \text{~ and }\\ 
111100, 101101, 100111, 110110, 010111, 011101, 110011, 111001 \ .
\end{array}\end{equation}
PALP presents the triangulations by {\it incidence} information; the interpretation is that each string of six 1's and 0's corresponds to a cone whose vertices are to be chosen from Eq.~\eqref{polytope24}, wherever a 1 is indicated in its position.
Indeed, there are precisely four 1's in each string, signifying that each cone is to have four generators, as required for a simplicial cone.
For completeness we present below some of the relevant output of PALP for this space, in response to the command 
\be
\nn \verb| poly.x -gPV| \ ,
\ee
and with the input being the weight system for $X_{2,4}$, 
\beq
\nn \verb|6 1 2 1 0 1 1  6 2 1 0 1 1 1| \; .
\eeq
This gives us the following screen output:
{\scriptsize
\begin{verbatim}
Degrees and weights  `d1 w11 w12 ... d2 w21 w22 ...'
  or `#lines #colums' (= `PolyDim #Points' or `#Points PolyDim'):
6 1 2 1 0 1 1  6 2 1 0 1 1 1
6 1 2 1 0 1 1  6 2 1 0 1 1 1 M:89 6 N:7 6 H:2,74 [-144]
4 7  points of P-dual and IP-simplices of codim >1 points
    0    3    0    0    1   -2    0
    0   -1    1    0   -1    1    0
    1    1    0    0    0   -1    0
    0    1    0    1    0   -1    0
------------------------------   #IP-simp=2  #prim.div.cl.=2
    1    1    0    1    1    2   6=d  codim=0
    1    0    1    1    2    1   6=d  codim=0
NewtonPoly=w+u^3wx/v+v+x+u/v+v/(u^2wx);  Pic=2
Incidence: 111100 101101 100111 110110 011111 111011
vp[0]=44 vp[1]=44 vp[2]=44 vp[3]=44 vp[4]=55 vp[5]=55
F4 2ndary dim=1 new circ.= C0 #maxTri=2
F5 2ndary dim=1 induced by C0
8 Triangulation
111100 101101 100111 110110 001111 011110 101011 111010
2 SR-ideal
010001 101110
SINGULAR -> divisor classes (integral basis J1 ... J2):
d1=J1, d2=J2, d3=J1-J2, d4=J1, d5=2*J1-J2, d6=J1+J2

SINGULAR -> intersection polynomial:
2*J1^3+1*J2^3+1*J1^2*J2-1*J2^2*J1 

SINGULAR -> Topological quantities of the CY-hypersurface:
c2(cy)= 14*J1^2+2*J1*J2-2*J2^2
Euler characteristic: -144

2 MORI GENERATORS / dim(cone)=2   [#rays=5<=16 #eq=2<=4 #v=2<=4]
  1 -1  2  1  3  0   I:10
  0  1 -1  0 -1  1   I:01
8 Triangulation
111100 101101 100111 110110 010111 011101 110011 111001
2 SR-ideal
001010 110101
SINGULAR -> divisor classes (integral basis J1 ... J2):
d1=J1, d2=J2, d3=J1-J2, d4=J1, d5=2*J1-J2, d6=J1+J2

SINGULAR -> intersection polynomial:
2*J1^3-5*J2^3+1*J1^2*J2-1*J2^2*J1 

SINGULAR -> Topological quantities of the CY-hypersurface:
c2(cy)= 14*J1^2+2*J1*J2-2*J2^2
Euler characteristic: -144

2 MORI GENERATORS / dim(cone)=2   [#rays=5<=16 #eq=2<=4 #v=2<=4]
  1  2 -1  1  0  3   I:10
  0 -1  1  0  1 -1   I:01
\end{verbatim}
}
The interpretation of the above output will be explained in the next Subsection, in relation to the geometrical data, especially to the computation of the K\"{a}hler and the Mori cones.

\subsection{K\"{a}hler and Mori cones}\label{KM}
All the geometrical data for the compact, smooth Calabi-Yau three-folds relevant to our heterotic models, can be easily read off from PALP. 
However, in order to determine the K\"{a}hler and Mori cones, special care is needed. Here, we explain how to interpret the PALP output by two examples. 

Detailed descriptions of the K\"ahler and Mori cones of $X$ are crucial in our vector bundle construction.
Let us choose a basis $\{ J_r \}$ for $r = 1, \ldots, h^{1,1}(X)$ of the K\"ahler forms of $X$ so that any general harmonic $(1,1)$-form $J$ can be expanded into this basis as $J = t^r J_r$ with parameters $t^r$.
One can represent the K\"ahler cone of $X$ by an $n_F \times h^{1,1}(X)$ integer matrix $K= \left[ K^{\bar{r}}_{~r}\right]$, such that $t^r$ are allowed K\"{a}hler parameters if
\beq \label{K-def}
K^{\bar{r}}_{~r} t^r \ge 0
\eeq
for all $\bar{r} = 1, \ldots, n_F$.
Here, $n_F$ is the number of facets of the K\"ahler cone which, clearly, cannot be less than the dimension of the cone, $h^{1,1}(X)$.
When the bound is saturated and $n_F = h^{1,1}(X)$ (i.e., whence the matrix $K$ is square) and when the generators are indeed linearly independent over $\IR$, the cone is called simplicial and the manifold {\em simple}.
In this case, the K\"ahler cone has exactly $h^{1,1}(X)$ generators, which we can denote as $\tilde{J}_r$, and which can then be set as the standard basis of $\IZ^{h^{1,1}(X)}$ by the linear transformations
\begin{equation} \label{newbasis}
\tilde{J}_r = (K^{-1})^t_{~r} J_t \ .
\end{equation}
Note that the matrix $K$ is square for a simplicial K\"ahler cone and hence we do not distinguish between the two types of index, omitting the bars in Eq.~\eqref{newbasis}.
We shall only study the simple manifolds whose K\"{a}hler cones are simplicial:
It turns out that 41 manifolds are non-simple amongst the total of 307 manifolds with $h^{1,1}(X)=3$, as described in \sref{2.1}, and that all manifolds with $h^{1,1}(X)=1,2$ are simple. 

The matrix $K$ is important in order to select the positive line bundles, where positivity is meant in the sense of the Kodaira vanishing theorem. Indeed, a line bundle $\cO_X(\bold a)$ is positive if \begin{equation}
K^{{r}}_{~s} a^s > 0 \quad \forall {r} = 1, \ldots, h^{1,1}(X) \ .
\end{equation}
Once the K\"{a}hler cone matrix $K$ is given we should next ask about the Mori cone.
Recall that the Mori cone is the cone of effective curve classes in $X$ and is 
dual to the K\"ahler cone, with respect to the pairing $\int_{C} {J} $ between curves $C$ and $(1,1)$ forms $J$.
PALP calculates the Mori cone data for us and presents the result in dot-product form.
Let us again illustrate this with the previous example, $X_{2,4}$, given in the end of \sref{palp-output}. 
Since $h^{1,1}(X_{2,4})=2$, we have two generators for the K\"ahler cone.
The dual polytope has six vertices; this means that there are six toric divisor classes to the toric variety.
Thus, for each of the two triangulations there is a re-writing of the six divisor classes $d_1$ to $d_6$ in terms of the divisor-class generators $J_1$ and $J_2$, as in the screen output.
We also see that the triple intersection numbers of $X$ are expressed as a cubic polynomial in $J_1$ and $J_2$, with the integer coefficients in front of the monomials denoting the intersection numbers for the corresponding triple products.
Shortly following is the second Chern class of $X$, also as a polynomial in $J_1$ and $J_2$.
Finally comes the two generators of the Mori cone, each corresponding to a row in the output. For example, 1 -1 2 1 3 0 in the first line of the first triangulation means that this generator dots into each of the six divisors $d_1$ to $d_6$, in the sense of the intersection product, to give these six integers. Again, translating back into the basis $J_1$, $J_2$ is straight-forward.

The upshot of the previous paragraph is that we can now obtain the K\"ahler and the Mori cones of the Calabi-Yau three-folds, from those of the triangulations through PALP.
We first emphasise that for each of the triangulations, the subsequent K\"ahler- and Mori-cone output suffices.
In the cases where MPCP-triangulations occur in multiple ways, we obtain a list of Mori cones $\cM_j$, one for each triangulation, and the Mori cone $\cM(X^p)$ of $X^{p}$ in the $p$-th {\it phase} 
is simply given as
\begin{equation} \label{mori-inter}
\cM(X^{p}) = \bigcap_{j {\text{ in phase }} p} \cM_j\ , 
\end{equation}
where the intersection is over all the triangulations in the given phase $p$. 
Recall that the {phase} was defined as a set of partially-desingularised toric fourfolds with the same intersection ring; from the definition, the number of phases for a given reflexive polytope can not be greater than the number of MPCP-triangulations.

Eq.~\eqref{mori-inter} means, by duality, that the K\"ahler cone $\cK(X^{p})$ of $X^{p}$ is the union
\begin{equation}
\cK(X^{p}) = \bigcup_{j {\text{ in phase }} p} \cK_j \ ,
\end{equation}
where $\cK_j$ is the K\"{a}hler cone for the $j$-th triangulation, and hence, is dual to $\cM_j$.

From the PALP output for $X_{2,4}$ given in the end of \sref{palp-output}, it is easy to see that the two partial desingularisations of $X_{2,4}$ give rise to two different intersection rings.
Therefore, we obtain the two Calabi-Yau three-folds, $X_{2,4}^1$ and $X_{2,4}^2$ corresponding to the two phases $p=1,2$, and no further complications arise. 

Let us now consider $X_{2,23}$ as another example. The Mori cone part of its PALP output gives the geometrical data of the following two MPCP-triangulations:

{\scriptsize
\begin{verbatim}
10 2 1 5 1 1 0  4 0 1 2 0 0 1
10 2 1 5 1 1 0  4 0 1 2 0 0 1 M:155 7 N:8 6 H:2,116 [-228]

9 Triangulation
110011 110110 111100 101110 111001 101011 001111 010111 011101
SINGULAR -> divisor classes (integral basis J1 ... J2):
d1=5*J1+2*J2, d2=J1, d3=J1, d4=J2, d5=2*J1, d6=J1+J2

SINGULAR -> intersection polynomial:
1*J2^3+1*J1^2*J2-1*J2^2*J1 

2 MORI GENERATORS / dim(cone)=2   [#rays=4<=18 #eq=2<=4 #v=2<=4]
  2  0  0  1  0  1   I:10
  0  2  2 -5  4 -3   I:01
  
8 Triangulation
110011 110110 111100 101110 111001 101011 011011 011110
SINGULAR -> divisor classes (integral basis J1 ... J2):
d1=5*J1+2*J2, d2=J1, d3=J1, d4=J2, d5=2*J1, d6=J1+J2

SINGULAR -> intersection polynomial:
1*J2^3+1*J1^2*J2-1*J2^2*J1 

2 MORI GENERATORS / dim(cone)=2   [#rays=4<=16 #eq=2<=4 #v=2<=4]
  3  1  1 -1  2  0   I:10
  0 -2 -2  5 -4  3   I:01
\end{verbatim}
}
As can be seen from the above PALP output, the two resolutions of $X_{2,23}$ have the same intersection ring. Therefore we are on the same manifold and the K\"ahler cone of the resulting space, $X^{1}_{2,23}$, is given by 
\begin{equation}
\cK(X^{1}_{2,23}) = \cK_1 \cup \cK_2 \ , 
\end{equation}
where $\cK_1$ and $\cK_2$ are the two K\"ahler cones of the two partially-desingularised ambient spaces. 
Since $d_2=J_1$ and $d_4=J_2$, the second and the fourth columns of the Mori data will give us the matrix of inner products $\int_{C_r} J_s$, where $C_r$ are the Mori cone generators and $J_s$ are the $(1,1)$-form basis elements. 
This matrix describes the facets of the K\"ahler cone and hence is exactly the matrix $K$ defined earlier. 
Reading off the two columns from the output, we obtain a K\"{a}hler-cone matrix for each of the triangulations
\begin{equation}
K_1= \left(
\begin{array}{ll}
 0 & ~~1 \\
 2 & -5 
\end{array}
\right) \ , ~
K_2= \left(
\begin{array}{ll}
~~1 & -1 \\
 -2 & ~~5 
\end{array}
\right) \ ,
\end{equation}
and by joining the two cones we obtain the K\"{a}hler-cone matrix for $X_{2,23}^1$:
\begin{equation}
K = \left(
\begin{array}{ll}
 0 & ~~1 \\
 1 & -1 
\end{array}
\right) \ .
\end{equation}
As for notation, $X_{h, i}^{p}$ denotes the toric Calabi-Yau three-fold obtained from the $i$-th reflexive four-polytope at $h^{1,1}(X)=h$, with the superscript meaning that it corresponds to the $p$-th phase of the desingularisations, due to multiple MPCP-triangulations. 
If a manifold is denoted by $X_{h,i}$ without an upper index $p$ it refers to the $i$-th four-polytope with a unique MPCP-triangulation. 

\section{Monad construction}
Turning to the gauge bundle construction, we shall search for monad bundles satisfying a certain set of constraints. 
The monad constructions of vector bundles has been developed in the mathematical literature~\cite{HM,beilinson,maruyama,monadbook} and, over the years, has been made extensive use of in the context of string model building~\cite{Distler:1987ee,Kachru:1995em,Blumenhagen:1997vt,Douglas:2004yv,maria,Blumenhagen:2006wj,Anderson:2008uw,Anderson:2009mh}. 
A monad $V$ over a Calabi-Yau three-fold $X$ is defined by the short exact sequence of the form
\be\label{monad}
0 \longrightarrow V  \longrightarrow \bigoplus\limits_{i=1}^{r_b} \cO_X(\bold{b}_i)  \stackrel{f}{\longrightarrow} \bigoplus\limits_{j=1}^{r_c} \cO_X(\bold{c}_j)  \longrightarrow 0 \ , 
\ee
where $\bold b_i$ and $\bold c_j$ are integer vectors of length $h^{1,1}(X)$, representing the first Chern classes of the summand line bundles $\cO_X(\bold{b}_i)$ and $\cO_X(\bold{c}_j)$. 
The bundle $V$ is a holomorphic $U(N)$-bundle, with its rank being 
$N=r_b- r_c$.

From Eq.~\eqref{monad}, one can readily read off the Chern class of $V$:
\bea
\label{c}
\nn c_1 (V) &=& \left( \sum_{i=1}^{r_b} b^r_i - \sum_{j=1}^{r_c} c^r_j \right) J_r \ , \\
c_2(V) &=& \frac{1}{2} d_{rst} \left(\sum_{j=1}^{r_c} c^s_j c^t_j - \sum_{i=1}^{r_b} b^s_i b^t_i \right) \nu^r \ , \\
\nn c_3(V) &=& \frac{1}{3} d_{rst} 
   \left(\sum_{i=1}^{r_b} b^r_i b^s_i b^t_i - \sum_{j=1}^{r_c} c^r_j
   c^s_j c^t_j \right) \ ,
\label{chernV}
\eea
where $J_r \in H^{1,1}(X, \IR)$ represent the harmonic $(1,1)$-forms $c_1(\cO_X(\bold e_r))$,  the $d_{rst}$ are the triple intersection numbers defined by
\beq \label{intersec-def}
d_{rst} = \int_X J_r \wedge J_s \wedge J_t \ , 
\eeq
and the $\nu^r$ are the 4-forms furnishing the dual basis to the K\"ahler generators $J_r$, subject to the duality relation
\beq \label{nubasis}
\int_{X} J_r \wedge \nu^s = \delta_r^s.
\eeq 
As can be seen from Eq.~\eqref{c}, the Chern class of $V$ only depends on the integer parameters $\bold b_i$ and $\bold c_j$, as well as the topology of the base manifold $X$.
Choosing an appropriate morphism $f$ in the defining sequence Eq.~\eqref{monad} corresponds to the tuning of more refined invariants of $V$. 

\subsection{Constraining the Bundles} 
The monad bundles should meet several mathematical and physical constraints which we shall describe shortly. However, we first recall that the new basis $\{\tilde{J}_r\}$ for $H^{1,1}(X, \IR)$ was obtained by the change of basis, Eq.~\eqref{newbasis}, starting from a rather arbitrary choice of the basis $\{J_r\}$. The K\"ahler cone has thereby become the positive orthant in $\IR^{h^{1,1}(X)}$.
The monad parameters in this new basis are denoted by ``tilded'' vectors $\tilde{\bold{b}}_i$ and $\tilde{\bold{c}}_j$, expressed in terms of the un-tilded in the obvious manner
\beq
\tilde {\bold{b}}^r = K^r_{~s} \bold{b}^s~; ~~\tilde {\bold{c}}^r = K^r_{~s} \bold{c}^s \ ,
\eeq
where $K^r_{~s}$ are the elements of the K\"ahler cone matrix $K$ defined in Eq.~\eqref{K-def}. 

%
\subsubsection{Positive Monads}\label{3.1.1}
As implied by the choice of terminology ``positive'' monads, we first restrict ourselves to those models with all monad parameters $\tilde{b}_i^r$ and $\tilde{c}_j^r$ positive. This is primarily a technical requirement which simplifies many calculations by Kodaira's vanishing theorem~\cite{HA}. It also has important phenomenological consequences.
For example, since $H^k(X, \cO_X(\tilde{\bold{b}}_i))=H^k(X, \cO_X(\tilde{\bold {c}}_j))=0$ for all $k>0$, it follows immediately that $H^2(X,V)=H^3(X,V)=0$ and hence, in particular, that no anti-families arise. 
There is also a more tenuous connection between positivity of monad parameters and stability of the vector bundle~\cite{Anderson:2007nc} but in this paper we do not address the stability issue. 

Secondly, we require that $\tilde c_j^r \geq \tilde b_i^r$ for all $r, i, j$ so that our monad actually becomes a bundle rather than a sheaf, provided that the morphism $f$ in Eq.~\eqref{monad} is sufficiently generic. 
Of course, to be free from redundancies, we assume that there does not appear a common line bundle piece in the two direct sums, that is, $\tilde{\bold b}_i \neq \tilde{\bold c}_j$ for all $i, j$. 

Finally, the Chern classes of our monad should also be constrained in such a way that $c_1(V)$ vanishes and $c_2(X) - c_2(V)$ lies in the Mori cone. 
The former comes from the ``special''-unitarity of the gauge bundle, and the latter from the cancellation of the heterotic anomaly. 
Upon applying Eq.~\eqref{c}, this leads to the following constraints on the integer parameters:
\bea
\nn \sum\limits_{i=1}^{r_b} \tilde{\bold{b}}_i &=& \sum\limits_{j=1}^{r_c} \tilde{\bold{c}}_j \ , \\ 
\nn \tilde{c}_{2}(TX)_r  &\geq&\frac{1}{2} \tilde{d}_{rst} \left[ \sum\limits_{j=1}^{r_c} \tilde{c}_j^s \tilde{c}_j^t  - \sum\limits_{i=1}^{r_b} \tilde {b}_i^s \tilde {b}_i^t \right] \ , \eea
where tilded quantities with lower $r,s,t$-type indices are obtained by transforming the un-tilded counterparts with $(K^{-1})^u_{~~r}$ as
\bea
\nn \tilde{d}_{rst} &=& d_{uvw} (K^{-1})^{u}_{~~r} (K^{-1})^{v}_{~~s} (K^{-1})^{w}_{~~t} \ , \\
\nn \tilde{c}_{2r}(TX) &=& (K^{-1})^{u}_{~~r} c_{2u}(TX)\ ,
\eea 
and sums over the repeated indices, from $1$ to $h^{1,1}(X)$, are implied if not explicitly written. 

Monad bundles that satisfy all the constraints described above will be called positive monads in this paper. 
Table~\ref{t:posmonadDef} summarises this set of criteria. 

\begin{table}[t!b!h!]
{\renewcommand{\arraystretch}{1.7}
{\begin{center} \small
~~~~~~~~\begin{tabular}{|c||c|c|} \hline 
\text{Origin} & \text{Conditions on the monad parameters} $\bold{\tilde{b}}_i$ \text{and} $\bold{\tilde{c}}_j$ \\[4pt] \hline \hline 
\text{Mathematical constraints} &  
$
\ba{rcl}
&(a).&\tilde{b}_i^r, \tilde{c}_j^r >0 \ , ~\forall r  \text{~~[positivity]~} \\
&(b).& \tilde{c}_j^r \geq \tilde{b}_i^r \ ,  ~\forall r  \text{~~[bundle-ness]~}\\ 
&(c).& \bold{\tilde b}_i \neq \bold{\tilde c}_j \text{~~[no redundancy]}
\ea 
$
\\[7pt] 
\hline

\text{~~Special-unitary gauge group~~} &  
$
\sum\limits_{i=1}^{r_b} \tilde {\bold b}_i = \sum\limits_{j=1}^{r_c} \tilde {\bold c}_j$
\\[7pt] 
\hline
{Anomaly cancellation} &
$
~~~~~\tilde{c}_{2}(TX)_r  \geq \frac{1}{2} \tilde{d}_{rst} \left[ \sum\limits_{j=1}^{r_c} \tilde{c}_j^s \tilde{c}_j^t  - \sum\limits_{i=1}^{r_b} \tilde {b}_i^s \tilde {b}_i^t \right] \ , ~\forall r
~~~~~$
\\[7pt] 
\hline
\end{tabular}
\end{center}}
\hspace{.85in}\parbox{6in}{\caption{\label{t:posmonadDef} \sf The list of criteria that a positive monad is defined to satisfy. 
}}\vskip -0.5cm
}
\end{table}

Before one attempts any systematic search for positive monads, a crucial piece of information is whether the scan is finite or not. 
A few lines of algebra (see Eq.~(5.7) in Ref.~\cite{Anderson:2008uw}) actually lead us to the following inequalities on $\tilde{b}_{max}^r \equiv \text{max}_i \{\tilde{b}_i^r\}$:
\beq \label{bmax}
\frac{2}{N}\tilde{c}_{2r} (TX) \geq  M_{rs} \tilde{b}_{max}^s\; . 
\eeq
Here, $N=r_b - r_c$ is the rank of $V$ and $M_{rs} = \sum\limits_{t=1}^{h^{1,1}(X)} \tilde{d}_{rst}$ . 
It turns out that these inequalities provide upper bounds for $\tilde{b}_{max}^r$ for every Calabi-Yau three-fold we consider in this paper. 
Having bounded the maximal entries in the first direct sum of Eq.~\eqref{monad}, we move on to finding an upper bound for the rank $r_b$. 
\comment{
Again by some algebra, we can get three independent ways to get upper bounds. 
Of course, there may be a more brilliant method than these three.
Once we get a bound for $r_B$, however, whether good or poor, that is already enough as long as computers can sort out all the remaining tasks concerning our classification programme. 
The classification has indeed finished and hence, we do not intend to go any further than these three methods. Below we list all the three without proofs, which can be found in \cite{cicy_posmonad}:
}
This once again proceeds along the same lines as in Section 5 of Ref.~\cite{Anderson:2008uw}.
There turn out to be three independent bounds, and for each Calabi-Yau we can check which one leads to the strongest constraint; the smallest upper bound is then used in any further calculations. These independent constraints on $r_b$ are inequalities (5.13), (5.14) and (5.16) of Ref.~\cite{Anderson:2008uw}:
\begin{enumerate}
\item
Given the calculated values of $\tilde{b}_{max}^r$, the following inequality gives us an upper bound: 
\beq \label{rb1}
r_b \leq N \left( 1+ \sum\limits_{r=1}^{h^{1,1}} \tilde{b}_{max}^r \right) \ . 
\eeq
\item
We first find non-negative integers $u^r$, satisfying 
\beq \label{rb2}
M_{rs} u^s \leq 2 \tilde{c}_{2} (TX)_r.  
\eeq 
Note that the inequality above has essentially the same form as the one~\eqref{bmax} for $\tilde{b}_{max}^r$ and hence, the solution space for the $u^r$ is finite. The non-negative integers $u^r$ are related to $r_b$ by 
\be \label{rb}
r_b = N+\sum\limits_{r=1}^{h^{1,1}} u^r  \ .
\ee
Given the finite solution set for $u^r$, we take the maximum of the corresponding $r_b$ values. 
\item
As in method 2, we first solve the inequality below for non-negative integers $u^r$: 
\beq \label{rb3} \sum\limits_{s=1}^{h^{1,1}} \left( 2 \sum\limits_{t=1}^{h^{1,1}} \tilde{d}_{rst} \tilde{b}_{max}^t + \tilde{d}_{rss} \right) u^s \leq 2 \tilde{c}_{2} (TX)_r + N \tilde{d}_{rst} \tilde{b}_{max}^s \tilde{b}_{max}^t. \eeq
Then we calculate all possible values of $r_b$ from Eq.~\eqref{rb} and find their maximum.
\end{enumerate}
Applying these inequalities not only allows us to prove the finiteness of the class, it also provides us with practical bounds on the entries $\tilde{\bf b}_i$, $\tilde{\bf c}_j$ and the ranks of the line bundle sums in the monad definition~\eqref{monad}. This allows us to set bounds for our scan which ensure that we indeed obtain an exhaustive list of models.
\subsubsection{Three Generations}
We also need to impose a basic three-generation constraint on our models. We require that the number of generations is a multiple of three, that is, ${\rm ind}(V)=-3 \kappa$ for $\kappa \in\mathbb{Z}_{\neq 0}$, and that some topological invariants of $X$, such as the Euler number, is divisible by the potential group order $\kappa$, that is, $\kappa \;|\;\chi(X)$. This is clearly necessary (although not sufficient) for the existence of a free quotient $\hat X = X/\Gamma$ with three generations ``downstairs'',  where $|\Gamma|=\kappa$. 

There exist a number of more refined topological invariants, given in Ref.~\cite{Candelas:1987du}, which can also be used to further constrain the order of potentially available free actions. These are the Euler characteristics $\chi({\cal N}^a\otimes TX^b)$ and Hirzebruch signatures $\sigma ({\cal N}^a\otimes TX^b)$ of the
``twisted'' bundles ${\cal N}^a\otimes TX^b$ (where ${\cal N}$ is the normal bundle of $X$) which must be
divisible by the group order $|\Gamma|$ for all integers $a,b\geq 0$. It was shown in Ref.~\cite{Candelas:1987du}, that it is sufficient to consider the cases $(a,b)=(0,1),(1,0),(2,0),(3,0)$ for the Euler characteristic and $(a,b)=(1,1)$ for the Hirzebruch signature without loosing information. We can compute these indices using the equations provided in Ref.~\cite{Candelas:1987du} and take their common divisors to obtain a list, $D(X)$, which must include the orders of all freely-acting symmetry groups for $X$.
Requiring that $\kappa$ is an element of this list, that is,
\beq \label{constraints}
\kappa=- {\rm ind}(V)/3 \in D(X) \ , 
\eeq 
can dramatically reduce the number of viable models.

\subsection{Classification of Positive Monads}
For the toric Calabi-Yau three-folds with $h^{1,1}(X) \leq 2$, the K\"{a}hler cones are simplicial. For this reason, a canonical (1,1)-form basis can always be chosen such that the K\"{a}hler cone corresponds to all the K\"{a}hler parameters being positive in the new basis.
Following Ref.~\cite{Anderson:2008uw, He:2009wi} by using the series of Eqs.~\eqref{bmax}--\eqref{rb3}, we conclude that the number of {{positive monads}} over the hypersurface Calabi-Yau three-folds with $h^{1,1}(X) \leq 2$ is finite. As for the Calabi-Yau manifolds with $h^{1,1}(X) =3$, the K\"{a}hler cones are not always simplicial, however, as stated earlier, we restrict ourselves to those with simplicial K\"ahler cones. Then, the same line of reasoning shows that the set of positive monad bundles over these manifolds is finite. In this Subsection we present the result of the scan of positive monads for each value of $h^{1,1}(X)$ in turn. 
\subsubsection{Bundles on $h^{1,1}(X)=1$ manifolds}
Dealing with the cyclic manifolds is relatively straightforward since the triangulations turn out to be unique in all five cases. Below, we summarise the relevant topological quantities requisite to the classification of the positive monad bundles. There is only one generator for the K\"ahler cone, denoted by $J$, so that $c_2(TX)$ is an integer multiple of $J^2$. We remark that $X_{1,2}$, the famous quintic, is the only one of the five manifolds with $h^{1,1}(X)=1$ that has a smooth ambient space; the other four all require (a single) desingularization of the ambient toric variety and thus these four were not considered in Ref.~\cite{He:2009wi}. The intersection matrix is a single number $d$ and we have also computed the twisted indices to find $D(X)$, the orders of possible freely acting symmetries. All these quantities are given in the table below.
\begin{equation}
{\renewcommand{\arraystretch}{1.2}
\begin{array}{|c|c|c|c|c|c|}
\hline
\mbox{Manifold} & h^{2,1} & c_2(TX)/J^2 & c_3(TX) = \chi & d & D(X)\\
\hline
X_{1,1} & 21  & 10&  -40 & 1 &  \{ 5 \} \\
X_{1,2} & 101 & 10& -200 & 5 &  \{ 5, 25 \}\\
X_{1,3} & 103 & 14& -204 & 3 &  \{ 3 \}\\
X_{1,4} & 145 & 34& -288 & 1 &  \{ \}\\
X_{1,5} & 149 & 22& -296 & 2 &  \{ 2, 4 \}\\
\hline
\end{array}
}
\end{equation}

We now proceed with the classification of positive monads.
On $X_{1,1}$, for example, we find 20, 14 and 9 positive monads of ranks 3, 4 and 5 respectively. 
The next manifold $X_{1,2}$ is the quintic, a cyclic CICY which was already analysed in \cite{Anderson:2007nc}. 
On this manifold, the same number of monads that have exactly the same integer parameters $\tilde b_i$ and $\tilde c_j$ as those on $X_{1,1}$, have been found with, of course, different Chern classes and different indices. 
Similarly, we also classify all the positive monads on the other  3 spaces. 

Table~\ref{t:pic1mon} summarises the resulting statistics of this bundle search, before and after imposing the index constraint, {{Eq.~\eqref{constraints}}}{, and also lists the potential group orders again}.

\begin{table}[h!t!]
\[
\begin{array}{|c|ccc|ccc|c|}
\hline
\mbox{Manifold} & 
\multicolumn{3}{|c|}{\# \mbox{(Positive Monads)}} & 
\multicolumn{3}{|c|}{\# \mbox{(Positive Monads with \eqref{constraints})}} &
\mbox{Potential group order} \\
\hline
       & ~~SU(3)~~ & ~~SU(4)~~ & ~~SU(5)~~ & ~~SU(3)~~ & ~~SU(4)~~ & ~~SU(5)~~ & D(X) \\
X_{1,1}& 20& 14 & 9 &  2 & 1 & 1 & \{5\}\\
X_{1,2}& 20& 14 & 9 & 3 & 1 & 1 &\{5,25\}\\
X_{1,3}& 53& 43 & 32& 1 & 0 & 0 &\{3\}\\
X_{1,4}& 1035 & 1182 & 1149 & 0 & 0 & 0 & \{\}\\
X_{1,5}& 230&218 &183 &  1& 1&0 & \{2,4\}\\
\hline
\end{array}
\]
\begin{center}
\parbox{6.5in}{\caption{{\em \sf { The statistics of positive monads on the 5 cyclic hypersurface Calabi-Yau three-folds $X$; the first two columns present the number of positive monads before and after the constraint \eqref{constraints} is imposed, repectively, and the last column, the set of potential group orders, $D(X)$.}}}
\label{t:pic1mon}}\end{center}\vskip -0.9cm
\end{table}
Note that, putting aside the phenomenologically less interesting $SU(3)$ monads, each of the two manifolds $X_{1,1}$ and $X_{1,2}$ admit a $SU(4)$ bundle and a $SU(5)$ bundle.  However, these two manifolds are the quintic quotiented by the (toric) $\IZ_5$-action and the quintic itself, where positive monads have already been {ruled out} completely \cite{Anderson:2009mh}. 
The only bundle left which catches the eye is, therefore, the single $SU(4)$ monad on $X_{1,5}$,  defined by the following short exact sequence
\begin{equation}
\sseq{V}{\cO(1)^{\oplus 7}}{\cO(2)^{\oplus 2} \oplus \cO(3)}{}{} \; .
\end{equation}
The index of this vector bundle is $-12$ and hence, we need either $\IZ_4$- or $\IZ_2 \times \IZ_2$-Wilson lines. 
However, neither of them can break $SO(10)$ down to a standard-like gauge group (see Appendix~\ref{B}). Therefore no realistic models from positive monads exist on the five cyclic hypersurface Calabi-Yau manifolds. 

\subsubsection{Bundles on $h^{1,1}(X)=2$ manifolds} \label{3.2.2} 
There are a total of 39 manifolds for $h^{1,1}(X)=2$ manifolds, nine of which have smooth ambient spaces and have already appeared in Ref.~\cite{He:2009wi}, while the remaining 30 require desingularization of the ambient toric variety. The relevant data, including Newton polynomials, for these manifolds can be found in \cite{link}. They fall into three categories, which we will denote as follows. Type (a) manifolds are those whose ambient space requires a single desigularization. Type (b) manifolds require multiple desingularizations each of which leads to a new manifold, which we will refer to as a {\it single-manifold phase}. Finally, type (c) manifolds require multiple desingularizations but with some giving isomorphic manifolds, the set of which we will refer to as a {\it multi-manifold phase}. We tabulate the number for each type in Table \ref{tt:h11=2}.
\begin{table}[h!t!] 
\small
\[
\begin{array}{|c|c|c||c|}
\hline
{\mbox{~~(a) Single desing.~~}} &
{\begin{array}{l}
$\text{~(b) Multi desing.}$ \\
$\text{[single-mfld.~phase]}$
\end{array}} &
{\begin{array}{l}
$\text{~(c) Multi desing.}$ \\
$\text{[multi-mfld.~phase]}$
\end{array}} & \text{total} \\
\hline
\hline
~24  & ~6 &  ~~~9  & 39 \\
\hline 
\end{array}
\]
\begin{center}
\parbox{6.5in}{\caption{\em \sf The number of manifolds with $h^{1,1}(X)=2$ for each of the three types: (a) Single desingularisation, (b) Multi desingularisation in a single-manifold phase, (c) Multi desingularisation in a multi-manifold phase.}
\label{tt:h11=2}}\end{center}\vskip -0.9cm
\end{table}

The search for three-generation models for non-cyclic base manifolds follows the same lines as the one for the cyclic manifolds. 
In Table \ref{t:pic2mon} we provide  the spaces which allow for $SU(4)$ or $SU(5)$ positive monads and we list the numbers of these bundles in each case; {it turns out that none of them satisfy the index constraint \eqref{constraints} as indicated by the zeros in the middle column of the table}. Note that we have not classified $SU(3)$ monads since they do not give rise to realistic models. {The symbols ``--'' in the table indicate that the bundle scan has not been undertaken; even if some positive monads are found to satisfy the index constraint~\eqref{constraints}, one cannot find appropriate discrete free actions in those cases (see Appendix~\ref{B}).}

\begin{table}[h!t!]
\[
\begin{array}{|c|cc|cc|c|}
\hline
\mbox{Manifold} & 
\multicolumn{2}{|c|}{\# \mbox{(Positive Monads)}} & 
\multicolumn{2}{|c|}{\# \mbox{(Positive Monads with {\eqref{constraints}})}} &
\mbox{Potential group order} \\
\hline
       & ~~SU(4)~~ & ~~SU(5)~~ &  ~~SU(4)~~ & ~~SU(5)~~ & D(X) \\
X_{2,1}& - & 14 & - & 0 & \{3\} \\ 
X_{2,5}& 43 & 14 & 0 & 0 & \{3,9\} \\ 
X_{2,9}& - & 56 & - & 0 & \{2,4\} \\ 
X_{2,13}& - & 19 & - & 0 & \{2,4\} \\ 
X_{2,28}& - & 5114 & - & 0 & \{2\} \\ 
X_{2,29}^1& - & 5114 & - & 0 & \{2\} \\ 
X_{2,29}^2 & - & 5114 & - & 0 & \{2\} \\ 
X_{2,36}& - & 736 & - & 0 & \{3\} \\ 
X_{2,23}^1 & - & 2298 & - & 0 & \{2\} \\ 
X_{2,30}^1 & - & 77146 & - & 0 & \{2\} \\ 
X_{2,32}^1 & - & 12962 & - & 0 & \{2\} \\ 
X_{2,33}^1 & - & 128 & - & 0 & \{2\} \\ 
X_{2,34}^1 & - & 2 & - & 0 & \{2\} \\ 
\hline
\end{array}
\]
\begin{center}
\parbox{6.5in}{\caption{{\em \sf { The statistics of SU(4) and SU(5) positive monads on the hypersurface Calabi-Yau three-folds $X$ with $h^{1,1}(X)=2$; the first two columns present the number of positive monads before and after the constraint \eqref{constraints} is imposed, repectively, and the last column, the set of potential group orders, $D(X)$. Only those spaces with non-zero number of bundles are tabulated. The symbols ``--'' indicate that the monads need not be considered according to the Wilson-line argument in Appendix~\ref{B}}.}}
\label{t:pic2mon}}\end{center}\vskip -0.9cm
\end{table}
\subsubsection{Bundles on $h^{1,1}(X) = 3$ manifolds} \label{3.2.3}
There are a total of 307 manifolds with $h^{1,1}(X)=3$,  28 of which have smooth ambient space and have appeared in Ref.~\cite{He:2009wi}, while the others requires desingularization of the ambient toric variety. The data for all these manifolds, including their Newton polynomials, can be found in \cite{link}. We group these manifolds into four categories: the first three types, (a), (b) and (c), are defined as for $h^{1,1}(X)=2$, and the last type (d), refers to manifolds that are not favourable. In this paper we do not intend to build heterotic bundles based on type (d) manifolds. Note that K\"ahler cones of dimension not less than 3 can be non-simplicial. There indeed arise 40 non-simple toric Calabi-Yau three-folds at $h^{1,1}(X)=3$. To avoid additional technical complications, they were ignored in the bundle scan. Table~\ref{tt:h11=3} shows the number of Calabi-Yau manifolds with $h^{1,1}(X)=3$ for each of the four aforementioned types. 
\begin{table}[h!t!] 
\footnotesize
\[
\begin{array}{|cc|cc|cc|c||cc|}
\hline
\multicolumn{2}{|c|}{\mbox{~~(a) Single desing.~~}} &
\multicolumn{2}{|c|}{\begin{array}{l}
$\text{~(b) Multi desing.}$ \\
$\text{[single-mfld. phase]}$
\end{array}} &
\multicolumn{2}{|c|}{\begin{array}{l}
$\text{~(c) Multi desing.}$ \\
$\text{[multi-mfld. phase]}$
\end{array}} &
\mbox{(d) Non-fav.} & \multicolumn{2}{|c|}{\mbox{~~Total}~~} \\
\hline
\hline
\mbox{~~~~~All} & \mbox{Simple} &\mbox{~~~~~All} & \mbox{Simple} &\mbox{~~~~~All} & \mbox{Simple} &\mbox{} & \mbox{~~All~~} & \mbox{Simple~} \\
~~~~~92 & [87] & ~~~~~84 & [79] & ~~~~~130 & [100] & 1 & 307 & [266] \\
\hline 
\end{array}
\]
\begin{center}
\parbox{6.5in}{\caption{\em \sf The number of manifolds with $h^{1,1}(X)=3$ for each of the four types: (a) Single desingularisation, (b) Multi desingularisation in a single-manifold phase, (c) Multi desingularisation in a multi-manifold phase, and (d) Non-favourable. The numbers in the square brackets only count the manifolds with simplicial K\"{a}hler cones, for which the bundle scan has been undertaken.}
\label{tt:h11=3}}\end{center}\vskip -0.8cm
\end{table}
Since we are neglecting non-simple and non-favourable manifolds, the strategy for scanning positive monads is very much the same as for the lower $h^{1,1}(X)$ cases. Table~\ref{t:pic3mon} presents the classification result of this scan. 
{Note that the entries for the possible symmetry orders, $D(X)$, are less than 9 for most of the base manifolds $X$ and hence, again for those cases, $SU(4)$ models are already ruled out by the Wilson-line argument (cf.~Appendix~\ref{B})}. 

\comment{
\begin{table}[h!t!]
\[
\begin{array}{|c|cc|c|}
\hline
\mbox{Manifold} & 
\multicolumn{2}{|c|}{\# ~\mbox{after constraint \eqref{constraints}}} &
{\mbox{Candidate group orders}} \\
\hline
& SU(4) & SU(5)  &D(X)  \\
{X_{3,244}^{2}}& - & {14}  &\{ 2 \} \\
\hline
\end{array}
\]
\caption{{\em The statistics of SU(4) and SU(5) positive monads over the hypersurface Calabi-Yau three-folds of multi-desingularisation type (single-manifold phase) with $h^{1,1}(X)=3$, after the constraint \eqref{constraints} is imposed, on top of the cancellation of anomaly. Of the 79 simple HCYs, only one manifold has a non-zero number of viable bundles, and hence, the others are not tabulated. The column on the right describes the set, $D(X)$, of possible group orders.}}
\label{t:MD-SM}
\end{table}}
\begin{table}[h!t!]
\[
\begin{array}{|c|cc|cc|c|}
\hline
\mbox{Manifold} & 
\multicolumn{2}{|c|}{\# \mbox{(Positive Monads)}} & 
\multicolumn{2}{|c|}{\# \mbox{(Positive Monads with \eqref{constraints})}} &
\mbox{Potential group order} \\
\hline
       & ~~SU(4)~~ & ~~SU(5)~~ &  ~~SU(4)~~ & ~~SU(5)~~ & D(X) \\
X_{3,11}^{1} & - & 660553 & - & 1044 & \{2,4\} \\ 
X_{3,104}^{1} & - & 10097 & - & 0 & \{3\} \\ 
X_{3,121}^{1} & - & 77778 & - & 0 & \{2,3,6\} \\ 
X_{3,148}^{1} & 65 & 0 & 0 & 0 & \{2,4,8,16,32\} \\ 
X_{3,182}^{1} & - & 515 & - & 0 & \{2,4\} \\ 
X_{3,205}^{1} & - & 660553 & - & 1044 & \{2,4,8\} \\ 
X_{3,206}^{1} & - & 7043 & - & 0 & \{2,4,8\} \\ 
X_{3,209}^{1} & - & 921 & - & 0 & \{2,4,8\} \\ 
X_{3,222}^{1} & - & 144 & - & 0 & \{2,4\} \\ 
X_{3,226}^{1} & 65 & 0 & 0 & 0 & \{2,4,8,16\} \\ 
X_{3,242}& - & 262 &  - & 0 &\{2\}\\ 
X_{3,243}& - & 210103 & - & 0 & \{2\}\\ 
X_{3,244}^2 & - & 5378 & - & 0 & \{2\} \\ 
\hline
\end{array}\]
\begin{center}
\parbox{6.5in}{\caption{{\em \sf {The statistics of SU(4) and SU(5) positive monads on the hypersurface Calabi-Yau three-folds $X$ with $h^{1,1}(X)=3$; the first two columns present the number of positive monads before and after the constraint \eqref{constraints} is imposed, repectively, and the last column, the set of potential group orders, $D(X)$. Only those spaces with non-zero number of bundles are tabulated. The symbols ``--'' indicate that the monads need not be considered according to the Wilson-line argument in Appendix~\ref{B}}.}}
\label{t:pic3mon}}\end{center}\vskip -0.9cm
\end{table}
We have the following observations from these results:
\begin{itemize}
\item { Upon imposing the index constraint~\eqref{constraints}, all but the two manifolds $X_{3,11}^1$ and $X_{3,205}^1$, both belonging to type (c), are ruled out. As for the bundle indices, all the monads on each manifold happen to have the same value, requiring the highest possible order in $D(X)$. For example, the $1044$ $SU(5)$ monads on $X_{3,11}^1$ have index $-12$ and would therefore need a free action of order four}.  

\item The numbers of viable bundles on $X_{3,11}^1$ and $X_{3,205}^1$ are the same. This is explained by the following observations. The intersection ring and the second Chern class of $X_{3,205}^1$ turn out to be twice those of $X_{3,11}^1$. It is clear from Table~\ref{t:posmonadDef} that the constraints for positive monads are invariant under the overall scaling of intersection numbers and second Chern class. Therefore, one obtains the same number of positive monads, each represented by the same integer parameters $\tilde{\bold b}_i$ and $\tilde{\bold c}_j$.
\end{itemize}
The full toric data for the {two} manifolds admitting viable monad bundles is shown in Table~\ref{t:pic3}.
\begin{table}[h!t!]
{\renewcommand{\arraystretch}{1.1}
\[
\begin{array}{|c|c|c|c|c|}
\hline
\mbox{Manifold} & h^{1,1} & h^{2,1} & \chi & \mbox{Newton Polynomial} \\
\hline \hline
X_{3,11} & 3&59&-112 & w u^2+\frac{z^2}{v^3 w^2}+v+\frac{z}{v^2
   w^2}+z+\frac{v w}{z}+\frac{z^2}{v^2 w^3 u^2} \\
X_{3,205} & 3&115&-224 & v u^2+\frac{v^2 u^2}{w}+\frac{v^2
   u^2}{z}+v+w+z+\frac{1}{v u} \\
\hline
\end{array}
\]}
\begin{center}
\parbox{6.5in}{\caption{{\em \sf The {two} $h^{1,1}(X)=3$ toric hypersurface Calabi-Yau three-folds which admit positive monads of potential phenomenological relevance. For each manifold, the Hodge numbers $h^{1,1}$ and $h^{2,1}$, the Euler number $\chi$, as well the Newton polynomial, are given.}}
\label{t:pic3}}\end{center}\vskip -0.9cm
\end{table}

\subsubsection{Summary}
\begin{figure}[t]
\centerline{
\includegraphics[trim=0mm 0mm 0mm 0mm, clip, width=7in]{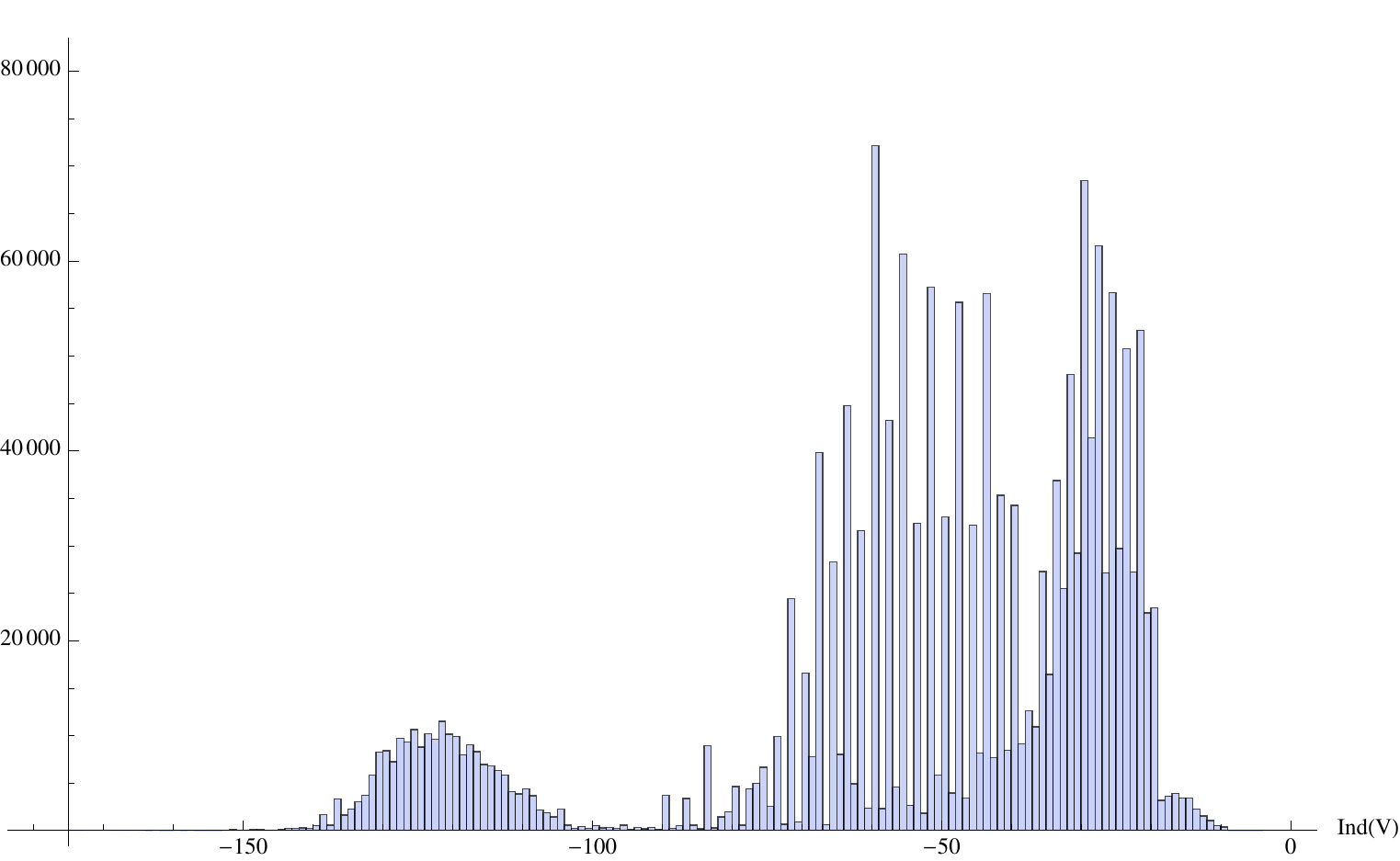}
}\vskip 0.2cm
\begin{center}
\parbox{6.5in}{\caption{{\sf { Histogram of ${\rm ind}(V)$ for the positive monad bundles $V$, of rank 4 or 5, over toric Calabi-Yau three-folds with Picard number 1, 2 and 3. The statistics of these bundles as well as the list of their base manifolds have been given in Tables~\ref{t:pic1mon},~\ref{t:pic2mon} and \ref{t:pic3mon}}.}}
\label{f:newplot}}\end{center}\vskip -0.9cm
\end{figure}

To summarise, {we have classified $1745090$ positive monads of rank $4$ or $5$, over $31$ toric Calabi-Yau three-folds. Amongst them, only $2088$ monads satisfy the family constraint~\eqref{constraints} on the index and also pass the group order test for Wilson-line breaking as explained in Appendix~\ref{B}. These bundles were found over only {2} hypersurface Calabi-Yau three-folds, { both with $h^{1,1}(X)=3$}. 

Figure~\ref{f:newplot} shows the distribution of the bundle indices for all the positive monads over the 31 base manifolds. 
It is natural that only a small fraction ($0.1\%$) survives the index constraint, given the fact that potential group orders tend to be less than or of order $10$. 
}


\subsection{Some example models}
Let us now present some explicit examples from the $2088$ bundles we have found. A comprehensive list of all these models is available at~\cite{link}. We have already seen that there are no phenomenologically interesting models based on positive monads for the five $h^{1,1}(X)=1$ manifolds and for the thirty-nine $h^{1,1}(X)=2$ manifolds. On the $h^{1,1}(X)=3$ manifolds, there is a total of $1044+1044$ models on the two base manifolds, $X_{3,11}^1$ and $X_{3,205}^1$. 
As we have observed in \sref{3.2.3}, each bundle on the two manifolds is represented by the same values of $\tilde{\bold b}_i$ and $\tilde{\bold c}_j$. 

One of the $1044$ ``admissible'' monads on $X_{3,11}^1$ is given by 
\begin{equation} \label{exmon}
\sseq{V}{\cO(1,1,1)^{\oplus 9}}{\cO(4,1,1) \oplus \cO(3,1,2) \oplus \cO(1,6,1) \oplus 
\cO(1,1,5)}{}{} \; .
\end{equation} 
This $SU(5)$ bundle has index $-12$ and requires a $\Gamma$-Wilson-line with $|\Gamma|=4$, in order to break the resulting $SU(5)$ GUT to the standard model gauge group. 
As shown in Table~\ref{t:pic3mon}, the indices of $X_{3,11}^1$ are indeed consistent with the existence of a freely-acting symmetry of order four.

The same short exact sequence, Eq.~\eqref{exmon}, defines another monad bundle on $X_{3,205}^1$, but now with index $-24$. Therefore, this monad requires a $\Gamma$-Wilson-line with $|\Gamma|=8$, which again is consistent with Table~\ref{t:pic3mon}.




\section{Conclusions and Prospects}
In this paper, we have systematically studied positive monad bundles on Calabi-Yau hypersurfaces in toric varieties with Picard numbers $h^{1,1}(X)\leq 3$, as a first step towards a systematic search for physical models on the largest known set of Calabi-Yau manifolds. First, we have shown that the class of such bundles, consistent with the heterotic anomaly condition, is finite and we have derived concrete bounds to facilitate exhaustive scans. In addition to imposing mathematical and physical consistency conditions, such as the heterotic anomaly condition, we have required two ``phenomenological" properties. First, the index of the bundle should equal $-3$ times the order of a freely-acting symmetry of the Calabi-Yau manifold. We have determined the possible orders of such freely-acting symmetries from topological invariants of the manifold. Secondly, for a given bundle the required symmetry order should allow for a Wilson line breaking of the GUT group to the standard model group. These two requirements turn out to be fairly restrictive. In total, we have found about {$2000$} bundles, all with $SU(5)$ structure group, on two manifolds, both with $h^{1,1}(X)=3$, satisfying these two physical requirements as well as the other consistency conditions. It is a promising sign that this number is significantly larger than that obtained, at a comparable scanning stage, for complete intersection Calabi-Yau manifolds~\cite{Anderson:2008uw}.

Before dividing the manifold by a discrete symmetry and including Wilson lines these bundles lead to $SU(5)$ GUT models with the net number of ${\bf 10}$ and $\bar{\bf 5}$ families equal to $3$ times a possible symmetry order. Since we are considering positive monads there are no $\bar{\bf 10}$ anti-families. In Ref.~\cite{Anderson:2008uw} it has been shown that positive monads on complete intersection Calabi-Yau manifolds have no vector-like ${\bf 5}$--$\bar{\bf 5}$ pairs and, hence, that, generically, one cannot obtain Higgs doublets. However, these arguments relied on properties of the projective ambient space, and it is not currently clear that they carry over to toric ambient spaces. To clarify this point more information on cohomology on toric varieties is required and the tools developed in Ref.~\cite{Blumenhagen:2010pv} will be helpful in this regard. Even if a generic Higgs turns out to be impossible, non-generic Higgs multiplets can arise, or be engineered, through specialization of the monad map $f$ in \eqref{monad}. 

Apart from the question of Higgs doublets, how far are our models away from quasi-realistic heterotic standard models? Of course, one has to show that our bundles preserve supersymmetry, that is, that they are stable $SU(5)$ bundles. In the context of complete intersection Calabi-Yau manifolds, positive monad bundles tend to be stable and the same might well be true in the toric case. However, in the absence of a general theorem stability would have to be checked explicitly, for example using the methods explained in Ref.~\cite{Anderson:2009nt}.
This requires significant information about bundle cohomology and is beyond the scope of the present paper. Another important question is whether a freely-acting symmetry of the required order indeed exists for at least some of our models. While divisibility of topological invariants of the manifold gives a good first indication of available symmetry orders, certainty can only be achieved by constructing the symmetries explicitly. 
Assuming a suitable symmetry indeed exists and vector-like ${\bf 5}$--$\bar{\bf 5}$ pairs are either generically present or can be engineered a further crucial condition is that all Higgs triplets can be removed from the spectrum while a pair of Higgs doublets can be kept. If this can be achieved the resulting spectrum will be that of the MSSM. In summary, we are three crucial model building steps away from quasi-realistic models. With {2000} bundles our data set is sizable and it is possible that these steps can be implemented for some of the models. This problem will be studied in a future publication.

\section*{Acknowledgments}
We would like to dedicate this paper to the memory of Maximilian Kreuzer. We are all indebted to him for his significant contributions, over the past decades, in creating the largest database of Calabi-Yau three-folds, and developing the necessary tools, conceptual and computational, to analyze them.

 A.~L.~is supported by the EC 6th Framework Programme MRTN-CT-2004-503369 and by the EPSRC network grant EP/l02784X/1. YHH would like to thank the Science and Technology Facilities Council, UK, for support through an Advanced Fellowship, the Chinese Ministry of Education, for a Chang-Jiang Chair Professorship at NanKai University, the US NSF for grant CCF-1048082, as well as City University, London and Merton College, Oxford, for their enduring support.

\begin{appendix}
\section{Wilson-line Breaking of $SO(10)$ and $SU(5)$}\label{B}
In Ref.~\cite{McInnes:1989rg}, the Hosotani mechanism which is used to break the GUT group to the standard model group via the inclusion of discrete Wilson line has been analyzed carefully, taking account of the group-theoretical subtleties. We can summarize the results relevant to our present study as follows: 

\begin{itemize} 
\item $SU(5)$ GUTs can be broken down to the Standard Model group $[U(1) 
\times SU(2) \times SU(3)] / \IZ_6$ 
by $\IZ_{n}$ for all $n \in \IZ_{>0}$ except $n=5$, 
via the explicit embedding of the Wilson line into $SU(5)$ as 
$\left( \begin{matrix} 
\gamma^3 I_{2 \times 2} & 0 \\ 0 & \gamma^{-2} I_{3 \times 3} 
\end{matrix} \right)$ with $\gamma \in \IZ_{n \ne 5}$. 

\item $SO(10)$ GUTs can be broken to the Standard Model group $SU(3) 
\times SU(2) \times U(1)^2$ by 
$\IZ_m \times \IZ_n$ where $m,n \in \IZ_{> 2}$, via the embedding 
of the Wilson line into $SU(4)$ as 
$\left( \begin{matrix} 
R(\theta) \otimes I_{2 \times 2} & 0 \\ 0 & R(\phi) \otimes I_{2 
\times 2} 
\end{matrix} \right)$ where $R(\theta) = \left( \begin{matrix} 
\cos \theta & -\sin\theta \\ \sin\theta & \cos\theta 
\end{matrix} \right)$. 
\end{itemize} 

Thus, for $SU(5)$ GUTs we need to exclude all symmetries of order $5$ 
while for $SO(10)$, we need to consider 
symmetries of at least order 9. 

\end{appendix}

\end{document}